\documentclass[twocolumn,english,prd,showpacs]{revtex4-1}
\usepackage[T1]{fontenc}
\usepackage[latin9]{inputenc}
\usepackage{verbatim}
\usepackage{graphicx}

\usepackage{babel}

\begin{document}

\title{Indirect Evidence for Lévy Walks in Squeeze Film Damping}

\author{S. Schlamminger, C.~A. Hagedorn, J.~H.~Gundlach}

\pacs{05.40.-a,07.10.Pz, 95.55.Ym,07.30.-t}

\affiliation{Center for Experimental Nuclear and Particle Astrophysics, University
of Washington, Seattle, WA 98195 USA}
\begin{abstract}
Molecular flow gas damping of mechanical motion in confined geometries,
and its associated noise, is important in a variety of fields, including
precision measurement, gravitational wave detection, and MEMS devices.
We used two torsion balance instruments to measure the strength and
distance-dependence of {}`squeeze film' damping. Measured quality
factors derived from free decay of oscillation are consistent with
gas particle superdiffusion in Lévy walks and inconsistent with those
expected from traditional Gaussian random walk particle motion. 
The distance-dependence of squeeze film damping observed in our experiments is in agreement with a parameter-free Monte Carlo simulation. The squeeze film damping of the motion of a plate  suspended a distance $d$ away from a parallel
surface scales with a fractional power between $d^{-1}$ and $d^{-2}$.
\end{abstract}
\maketitle
The damping of mechanical motion by residual gas particles in vacuum
has been the subject of measurement, discussion, and controversy~\cite{Li - Free Damping,Kadar - Free Damping,Zook -Free Damping,Christian - Free Damping,Quartz Fiber Paper}.
`Squeeze film damping' is enhanced gas damping that occurs due to
molecular flow in constrained geometries~\cite{Suijlen,Italian PRL,Bao - Free Damping,Hutcherson- Free Damping}.
Quantitative understanding of this damping is essential to a wide
range of subjects: The quality factor of many MEMS oscillators is
limited by squeeze film effects. In future gravitational wave interferometers,
test masses will be suspended within millimeters of another surface,
constraining gas flow. The fluctuation-dissipation theorem dictates
that squeeze film damping generates noise that may limit these experiments~\cite{Saulson}.

Noteworthy are two recent publications on squeeze film damping: In Ref.~\cite{Suijlen} the damping of three topologically different MEMS oscillators suspended a fixed distance from a substrate was measured as function of residual gas pressure. Ref.~\cite{Italian PRL} also measured the damping as function of gas pressure, using two torsion balances centered in closely-spaced housings. Both groups measured greater damping than expected from free gas damping, and developed Monte Carlo simulations that agreed with their experimental data. Ref.~\cite{Italian PRL} drew attention to the importance of constrained gas damping to gravitational wave detectors. As their geometry models the planned LISA gravitational wave detector, which is sensitive to both transverse and longitudinal gas flow, they did not replicate the planar geometry of existing and planned ground-based interferometers. These experiments and the failure of analytic theory to match their results motivated our measurement of the distance dependence of squeeze film damping with a simple planar geometry.

We used two different torsion balances to investigate mechanical damping
in rarefied gas. We provide experimental data characterizing the distance
and pressure dependence of squeeze film damping using the simple and
well controlled geometries of two adjacent surfaces. We compare our
data to analytic theory and to Monte Carlo simulations. The simulations
indicate that Gaussian random walks incorrectly describe the diffusion
of gas particles in narrow gaps. Instead, squeeze film damping, including
distance-dependence, is better described by superdiffusive particle
transport in Lévy walks.

Each of our two torsion balance instruments consists primarily of
a thin vertical plate pendulum suspended from a torsion fiber. Each pendulum is parallel to a stationary surface situated a distance 
\emph{$d$} away. We determined the damping by measuring the mechanical
quality factor, $Q$, of each pendulum's free oscillation as a function
of \emph{$d$} and as a function of the residual gas pressure $P$. 

\begin{figure}
\includegraphics[width=0.8\columnwidth]{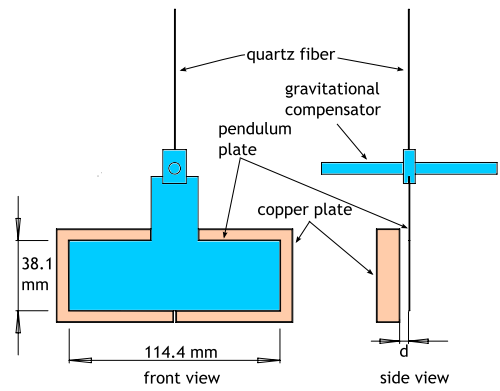}

\caption{Schematic drawing of the LTA torsion pendulum. The copper plate is
mounted on a translation stage allowing gaps up to $9$~mm. The
pendulum is suspended by a quartz fiber with $Q>50000$.\label{fig:LISA} }

\end{figure}

The LISA Test Apparatus (LTA) torsion balance~\cite{LisaSetup} depicted
in Fig.\ \ref{fig:LISA} was constructed specifically to investigate
noise sources in LISA. 
The plate of the LTA torsion pendulum is a 0.45 mm thick Si wafer,
$114.4$~mm wide and $38.1$~mm tall. The plate has a small riser
section connecting it to a perpendicular Al bar designed to minimize
its sensitivity to fluctuating gravitational fields caused by nearby
human activity. The pendulum assembly is suspended from a quartz fiber
with intrinsic $Q>50000$. The assembly has a free period of $120.89$~s
and a moment of inertia of $9.7\times10^{-6}$~kg$\,$m$^{2}$. The pendulum
plate is positioned parallel to a thick Cu plate. The Cu plate is
divided into two halves; both are electrically grounded. The pendulum
to Cu-plate gap, $d$, can be varied between 0 and 9 mm. The Cu plate
and pendulum are sputter-coated with Au to reduce contact potentials
and patch effects~\cite{SurfPot Paper}. The total electrical charge
on the pendulum was reduced below 0.5~pC~\cite{Charging Paper};
our results were not affected by deliberate increases in charge. The
torsion balance is housed within a stainless-steel vacuum chamber.
The vacuum is maintained by a small turbo-molecular pump with an extra
molecular drag stage backed by a small diaphragm pump. 

\begin{figure}
\includegraphics[clip,width=0.4\textwidth]{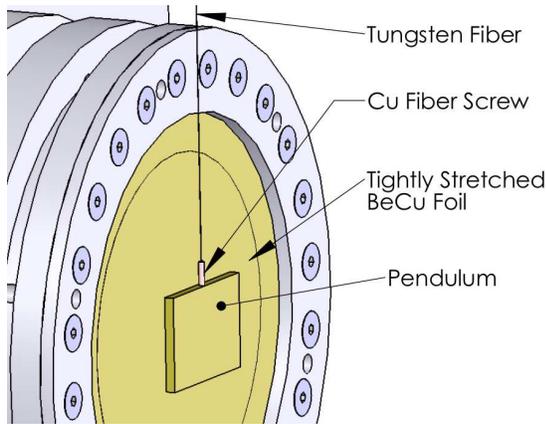}

\caption{Schematic drawing of the NTA torsion pendulum. The pendulum is suspended
by a tungsten fiber with intrinsic $Q\sim4000$. The pendulum can
be translated to change the gap between pendulum and foil.\label{fig:ISL}}

\end{figure}

Our second instrument, the `Newton Test Apparatus' (NTA), is depicted
in Fig.\ \ref{fig:ISL}. A $2.00$ mm thick, $42.18$ mm wide, and
$30.81$ mm tall metal plate is suspended from a tungsten fiber near
a tightly-stretched $12$~$\mu$m thick flat beryllium-copper foil.
The foil-pendulum gap, $d$, is varied by moving the torsion fiber
suspension point. The pendulum has a free period of $85$~s and a
moment of inertia of $3.1\times10^{-6}$~kg$\,$m$^{2}$. Like in the
LTA, both foil and pendulum are coated with Au. The torsion balance
is housed within a stainless-steel vacuum bell jar and the vacuum
is maintained by a turbo-molecular pump.

The torsional motion of each pendulum is monitored with a laser autocollimator.

To determine the size of the gap and to assure proper alignment between
the pendulum plates and their adjacent surfaces in LTA and NTA, several
methods were employed: (1) Calibrated photographs were taken through
vacuum windows. (2) Each pendulum was allowed to twist in both directions
until the pendulum's outer edge touched the adjacent surface. The
gap was determined by the maximum twist angles and the size of the
pendulum plate. (3) In the LTA, the capacitance between the two halves
of the copper plate was measured as a function of $d$ and was fit
to a capacitance model. Discrepancies between the distance measurements
determined the uncertainty $\sigma_{d}=0.1$~mm in LTA, $15$~$\mu$m
in NTA\@. The parallelism, about the horizontal axis, of the pendulums
w.r.t the adjacent surfaces was better than $0.5$~mrad in LTA and
$30$~$\mu$rad in NTA\@. The parallelism about the vertical axis
was better than $0.15$~mrad in LTA and 0.6~mrad in NTA. Systematic
and flatness distance uncertainties were $<45$~$\mu$m in LTA and
$<23$~$\mu$m in NTA\@.

To improve confidence in our pressure measurement, we connected five
ion gauges from three different manufacturers and a variable leak
valve to a single small vacuum chamber. With the exception of one
suspect gauge, the gauges agreed to within 30\% over the range of
$10^{-4}$--$2.6\times10^{-2}$~Pa. From this set, a cold-cathode
gauge was used for all of the LTA measurements and a hot cathode gauge
was used for all of the NTA measurements. We calibrated the hot-cathode
ion gauge with a simple pressure-division apparatus. With it, we verified
the gauge linearity and calibrated the gauge absolutely to within
$2\times10^{-3}$~Pa. Our experiments were carried out at a temperature
of $297\pm1$~K.

In the LTA, the pressure was adjusted by changing the speed of the
turbo pump and allowing the apparatus to come into equilibrium. Data
were taken at three different pressures. The pressure at the gauge
was $\sim10\%$ smaller than the pressure in the chamber due to gas
pumping impedances. We measured the composition of the gas in the
chamber with a residual gas analyzer at all three working pressures.
H$_{2}$O ($18$~amu) and N$_{2}$ ($28$~amu) together accounted
for at least $85$\% of the residual gas. At the working pressures
of $2\times10^{-3}$, $1\times10^{-4}$, and $4\times10^{-5}$~Pa,
the partial pressure ratios ($p_{18}$:$p_{28}$) were 0.4:1, 2:1,
and 8:1, respectively. Nitrogen is pumped more efficiently at high
turbo pump speeds. 

In the NTA, measurements were made by pumping the vacuum vessel below
$10^{-4}$~Pa, closing the turbo pump's roughing valve, and turning
the turbo pump off. $Q$ measurements began after the turbo had stopped
($P\geq9\times10^{-4}$~Pa). The pressure in the chamber rose at
a constant rate, most likely due to outgassing from the chamber walls.
Data were taken with pressure rises of $1.3$--$7\times10^{-7}$~Pa/s
until the pressure reached $1.3$--$2\times10^{-2}$~Pa. We assume
that the residual gas was primarily particles with mass 18~amu, as
we expect the outgassing to be dominated by water. 

We measured the free decay of each pendulum's torsional mode to determine
$Q$. The LTA torsion pendulum was excited to an oscillation amplitude
of $0.9$~mrad. The pendulum motion was recorded for $>10000$~s.
The data were cut into segments two oscillation periods in length.
The data in each cut were fit to $\theta(t)=A\cos(2\pi ft)+B\sin(2\pi ft)+C$.
The extracted amplitudes were then fit to an exponential decay to
extract $Q$, where the amplitude $\theta(t)=\theta_{0}e^{-\pi ft/Q}=\sqrt{A(t)^{2}+B(t)^{2}}$
. Errors of 20\% were deemed to be conservative overestimates in all
cases. In the NTA, data recordings of >1000~s, with initial amplitude
$0.2$--$4$~mrad, were fit to $\theta(t)=e^{-\pi t/TQ}\left(A\cos(2\pi t/T)+B\sin(2\pi t/T)\right)+C+Dt$,
where $A,B,C,D,T$, and $Q$ were fit simultaneously. This procedure
allowed measurements while the gas pressure was slowly varied. Seismic
noise and other excitations conservatively limit our ability to measure
$Q$ to $<50000$ in the LTA and $<1500$ in the NTA\@. 

\begin{figure}
\includegraphics[width=1\columnwidth]{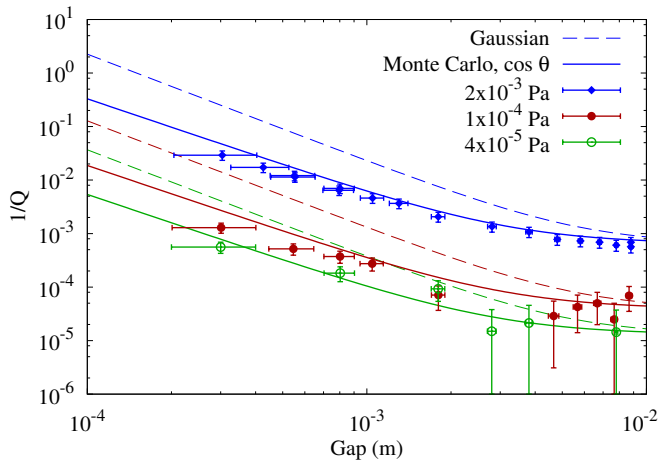}

\caption{The dissipation factor ($1/Q$) as a function of plate-pendulum separation
at constant pressure in the LTA\@. The lines are calculated using
Eq.~\ref{eq:FullDamping} with the diffusion time $\tau$ determined
by the parameter-free Monte Carlo simulation (solid) and the Gaussian
diffusion equation (dashed). The curves flatten with increasing gap
as unconstrained gas damping begins to dominate. The residual gas
composition changes with pressure; the curves are calculated using
the five measured leading gas species at each pressure. \label{fig:LISADamping}}

\end{figure}

\begin{figure}
\includegraphics[angle=270,width=1\columnwidth]{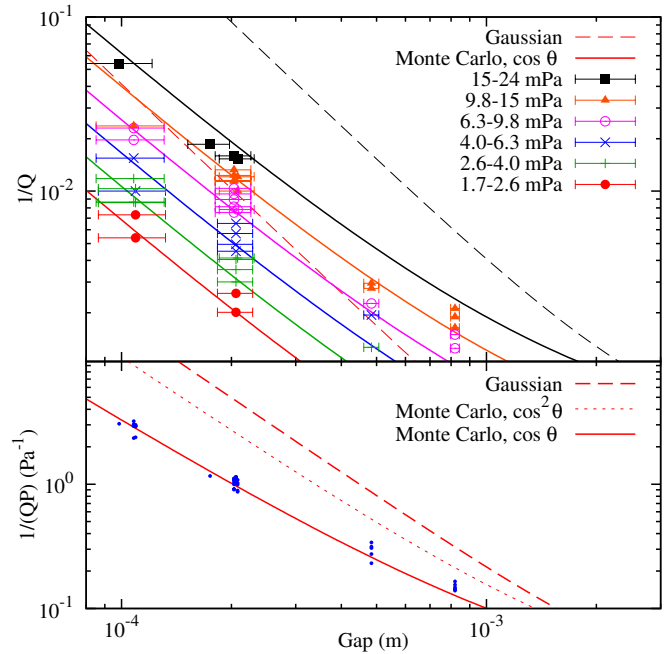}

\caption{Dissipation factor as a function of pendulum-foil separation in the
NTA\@. The lines are calculated as in Fig.\ \ref{fig:LISADamping}.
\label{fig:ISLDamping} The lower panel shows the dissipation factor
divided by the pressure. These data agree well with our Monte Carlo
prediction (solid line), which has superdiffusive Lévy-like behavior,
The data are inconsistent with Gaussian theory (dashed line) and with
a desorption angular distribution $p(\theta)\propto\cos^{n}\theta$,
for $n\geq2$ (dotted line).}

\end{figure}

The inverse of the measured quality factors, as functions of pressure
and pendulum-plate separation, are presented in Figs.\ \ref{fig:LISADamping}
and \ref{fig:ISLDamping}.

In the molecular flow regime, where the particle mean free path is
much larger than the gap and particle interactions may be neglected,
gas damping is primarily due to `unconstrained gas damping' and `squeeze
film damping'. All of our data were taken in this regime.

In unconstrained gas damping, the motion of an object is retarded
by collisions with ambient gas particles. A rectangular plate of dimensions
$a\times b$ moving perpendicular to its surface with velocity $u$
through a dilute gas is retarded by the force: $F(u)=-\alpha P ab\langle v\rangle^{-1}u$,
with $\langle v\rangle=\sqrt{8k_{B}T/(\pi m)}$, $m$ the gas particle
mass, the ambient pressure $P$ and temperature $T$. Three values
for $\alpha$ can be found in the literature: $\alpha=16/\pi$~\cite{Christian - Free Damping,Hutcherson- Free Damping},
$\alpha=16$~\cite{Kadar - Free Damping}, and $\alpha=24$~\cite{Li - Free Damping}.
Unconstrained gas damping depends linearly on the gas pressure but
is independent of the gap to nearby surfaces. 

Squeeze film damping of two adjacent plates moving towards each other
arises because the gas between the plates is temporarily compressed
until the {}``trapped'' gas can diffuse out of the gap and restore
equilibrium with the surrounding gas. To compute the dissipative force
on the plate we start with Suijlen's derivation~\cite{Suijlen}.
Periodic motion of the plate with amplitude $z_{0}\ll d$ and angular
frequency $\omega\ll\langle v\rangle/d$ will cause the distance-dependent
squeeze film force on the moving plate:

\begin{equation}
F=-\frac{abP}{d}\frac{(\omega\tau)^{2}+i\omega\tau}{1+\omega^{2}\tau^{2}}z_{0}\mbox{,}\label{eq:DampForce}\end{equation}
 where the diffusion time, $\tau$, for particles to enter or leave
the gap is discussed in detail below.

For a torsional oscillator, the damping forces must be converted to
torques. We approximate the torsion pendulum as two $a\times b$ plates
with lever arm $a/2$. We expect that a more precise model would give
only small corrections to this approximation. For a torsion balance
with resonant frequency $f_{0}\ll1/\tau$ and moment of inertia $I$,
we find that the inverse of the quality factor $Q$ is:

\begin{equation}
\frac{1}{Q}=\frac{a^{3}bP}{4\pi If_{0}\langle v\rangle}\left(\alpha+\frac{\langle v\rangle\tau}{d}\right)\mbox{.}\label{eq:FullDamping}\end{equation}

Since Suijlen \emph{et al.}\ \cite{Suijlen} and Cavalleri \emph{et
al.}\ \cite{Italian PRL}\emph{ }found that their experimental results
and simulation disagreed with Gaussian theoretical expectations, we
undertook our own simulation effort to determine the diffusion time
$\tau$. We simulated gas particles traveling interaction-free in
the gap. Initially, particles were randomly distributed throughout
the gap volume and assigned three-dimensional velocities according
to the Maxwell-Boltzmann distribution. Upon collision with a wall,
the particles were emitted with a polar angular distribution $p(\theta)$
and a Maxwellian velocity distribution. The angular and velocity distributions
of desorbed particles are poorly understood. A review,~\cite{DesorptionReview},
suggests that $p(\theta)\propto\cos^{n}\theta$ with $n\geq1$, where
the traditional choice, $n=1$, is motivated by time-reversal symmetry.
The particles were tracked until they left the volume, and the time
at which they left the gap was recorded. The average of these times,
for 10000 trials, was taken as the simulated diffusion time. For the
setup for which Suijlen \emph{et al.} provide a complete description
of their geometry, our results agree. 

\begin{figure}
\includegraphics[angle=270,width=1\columnwidth]{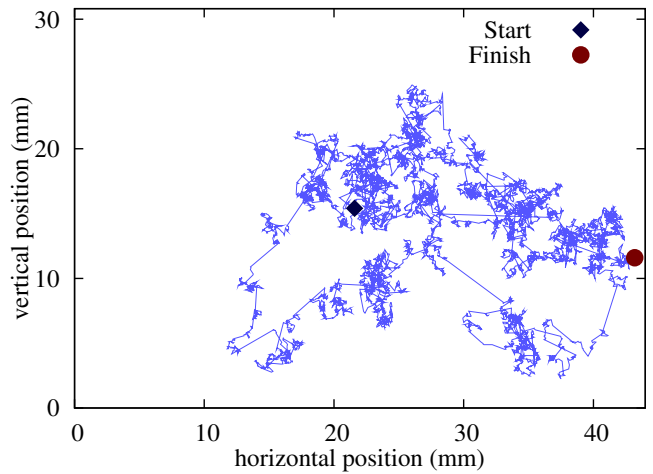}

\caption{Simulated trajectory of a gas particle in the NTA experiment for a
$0.1$~mm gap. The occasional large steps are a consequence of the
infinite variance of the step-size distribution. These steps, which
lead to superdiffusion, and scale-invariance are characteristic of
Lévy walks.\label{fig:LevyWalk}}

\end{figure}

Simulated particle tracks (Fig.\ \ref{fig:LevyWalk}) show occasional
very long steps. These long steps are essential for understanding
squeeze film damping.

Assuming thermal equilibrium and $p(\theta)\propto\cos\theta$, the
probability for a particle to make a step of length $r_s$ with a duration
 $t_s$ is given by \begin{equation}
p_{1}(r_s,t_s)\mbox{d}r_s\mbox{d}t_s=\frac{64d^{2}r_s\exp{(-\frac{4(d^{2}+r_s^{2})}{\pi t_s^{2}\langle v\rangle^{2}})}}{\pi^{2}t_s^{4}\langle v\rangle^{3}(d^{2}+r_s^{2})^{1/2}}\mbox{d}r_s\mbox{d}t_s.\label{eq:LevyDist}\end{equation}
 The length of the step, \emph{$r_s$}, is measured parallel to the
plates. $r_s$ and $t_s$ are not independent, but the distribution of
each can be obtained by integrating $p_{1}(r_s,t_s)$ over the other.
For example, the steps are distributed as $p_{1}(r_s)\mbox{d}r=2d^{2}r_s/(d^{2}+r_s^{2})^{2}\mbox{d}r_s$,
consistent with~\cite{Suijlen}. The step-size and time-of-flight
distributions have infinite variance, indicative of Lévy flights/walks.
In a Lévy flight the time to make a step is independent of step length.
In a Lévy walk the particle travels with finite velocity~\cite{Shl93,Kla96 Physics Today,Shl99,Bar02},
as in a physical gas. The defining feature of Lévy flights/walks is that repeated
application of the step-size distribution leads to a stable, scale-invariant
distribution with infinite variance. The Lorentzian is an example
of such a distribution. Our Monte Carlo simulations show that $p_{N}(r)$
becomes scale invariant after a few bounces. Lévy walks can lead to superdiffusive
behavior~\cite{Kla96 Physics Today,Bar02}, i.e.\,the mean square displacement
of an ensemble of particles released at $r=0$ increases as $\langle r_{N}^{2}\rangle\propto t^{\gamma}$,
with $\gamma>1$. Our Monte Carlo simulation of gas particle trajectories
trapped between two infinite planes shows $\gamma\approx1.15$, independent
of $d$, for $t\gg d/\langle v\rangle$, i.e.\ after a few bounces.
$\langle r_{N}^{2}\rangle$, in contrast to $\langle r_s^{2}\rangle$,  remains finite because  $\langle r_{N}^{2}\rangle$ is evaluated using the position of all particles in free flight between collisions with the two planes.


The smooth curves in figures \ref{fig:LISADamping} and \ref{fig:ISLDamping}
are computed using Eq.\ \ref{eq:FullDamping}, with $\alpha=16$.
The dashed lines are calculated with the Gaussian diffusion time,\emph{
}$\tau=8ab/(\pi^{3}d\langle v\rangle)$~\cite{Suijlen}, and the
solid lines are computed using our Monte Carlo diffusion times. The
dotted line in Fig.\ \ref{fig:ISLDamping} is computed using the
same simulation but with an angular distribution of emitted particles
$p(\theta)\propto\cos^{2}\theta$. As the damping is expected to be
linear in pressure, we also plot $1/(QP)$ for the NTA\@. We cannot
do so for the LTA results, as the residual gas species change with
pressure. The lines in Fig.\ \ref{fig:LISADamping} are calculated
with the measured residual gas composition.

Our squeeze film measurements are inconsistent with Gaussian diffusion.
In Eq.~\ref{eq:FullDamping}, for small $d$, $Q$ is proportional to $d^\beta$, where $\beta$ depends on the diffusion time, $\tau(d)$. For $\tau\propto d^{-1}$ (Gaussian diffusion), $\beta=2$  and for constant $\tau$,  $\beta=1$.  We find $\beta=1.4\pm0.3$ in LTA and $1.6\pm0.3$ in NTA\@. Our Monte Carlo
simulations predict $\beta=1.73$ in LTA and $1.85$ in NTA\@. The
data are consistent with our simulations using $p(\theta)\propto\cos^{n}\theta$
for\emph{ $n=1$} and clearly inconsistent for $n\geq2$. The measured
$Q$ is inversely proportional to gas pressure, as predicted in Eq.\ \ref{eq:FullDamping}.



We have made the first rigorous measurement of the distance dependence
of squeeze film damping by adjusting the gap between an oscillator and a flat plane. 
Experiments in two separate instruments were performed using planar geometries such that the gas flow is restricted by a single gap. This geometry is analogous to the geometry in existing gravitational wave interferometers and simpler than the geometry of the experiments decribed in  Ref.~\cite{Italian PRL}. 
We point to the connection between squeeze film damping
and the theory of Lévy walks. The infinite variance of the step-size
distribution of Lévy walks leads to superdiffusion, which is consistent
with our data. These measurements are simple examples of microscopic
Lévy phenomena in nature. In addition, our results support a cosine
angular distribution of re-emission of adsorbed gas particles. As
predicted by our Monte Carlo simulation, the measured squeeze film
damping is smaller and less dependent on gap size than expected by
Gaussian theory. The comprehensive data set presented here can be
used to verify any Monte Carlo simulation of squeeze film damping.
Squeeze film damping must be considered as a limitation of next-generation
gravitational wave detectors. The design of these detectors requires
precise knowledge of the strength and distance dependence of squeeze
film damping.

We thank Rainer Weiss for stimulating conversation. This work was
supported by NASA grant NNX08AY66G, by NSF grant PHY0653863, and by
DOE funding for the CENPA laboratory. This article has been assigned LIGO document number LIGO-P1000013.

\end{document}